\documentclass[conference]{IEEEtran}
\IEEEoverridecommandlockouts
\usepackage{cite}
\usepackage{amsmath,amssymb,amsfonts}
\usepackage{algorithmic}
\usepackage{graphicx}
\usepackage{textcomp}
\usepackage{xcolor}
\usepackage{booktabs, multirow,array}
\usepackage{float}
\usepackage{subfig}
\usepackage{upgreek}

\def\BibTeX{{\rm B\kern-.05em{\sc i\kern-.025em b}\kern-.08em
    T\kern-.1667em\lower.7ex\hbox{E}\kern-.125emX}}
\begin{document}

\title{A Material Sensing-Assisted Initial Beam Establishment Method for JCAS Systems}

\author{\IEEEauthorblockN{Yi Geng}
	\IEEEauthorblockA{\textit{Cictmobile, China} \\
		gengyi@cictmobile.com}
}

\maketitle

\begin{abstract}
Communication systems operating at high frequency bands must use narrow beams to compensate the high path loss. However, it is incredibly time-consuming to achieve beam alignment between the transmitter and receiver due to the large volume of beam space with narrow beams. The high latency of initial beam establishment will challenge the implementation of future 6G networks at high frequency bands. To tackle this problem, this paper proposes an initial beam establishment method using the material sensing results from joint communications and sensing (JCAS) systems. The reflection loss (RL) induced by each reflector can be predicted by exploiting the pre-identified material information of reflectors in the environment. The base station (BS) first scans the beam directions with low RL and establishes the connection immediately without sweeping the rest of the beam directions. In this way, the latency of initial beam establishment is significantly reduced.

\end{abstract}


\section{Introduction}
In the timeframe of 2030, 6G is expected to increase the communication capacity by a factor of 20 than 5G\cite{nokia}. One of the key enablers for this improvement is higher frequency bands (e.g., mmWave or even THz), where much higher bandwidths can be utilized. However, using higher frequency bands provides many challenges, such as high atmospheric attenuation\cite{8732419}. To compensate the high path loss, extraordinarily narrow beams must be used. However, the use of narrow beams would pose a significant challenge when performing initial beam establishment. Initial beam establishment aims to establish a suitable beam pair, including a base station (BS)-side beam direction and a user equipment (UE)-side beam direction that jointly provides good connectivity between the BS and UE. In 5G NR, multiple synchronization signal blocks (SSB) are exhaustively swept in all beam directions in the coverage area. The UEs measure the reference signal received power (RSRP) of SSBs in all beam directions. If the RSRP in a beam direction exceeds a threshold, the network will establish the connection in this beam direction. However, this procedure becomes more challenging at high frequency bands. The exhaustive beam search with narrow beams would result in a large number of sweeps. As an example, consider an antenna array with a 3~dB beamwidth of $5^{\circ}$ and field of view (FOV) of $\pm70^{\circ}$ both in azimuth and elevation. Taking a exhaustive beam search needs to perform 784 sweeps, which leads to a high latency of initial beam establishment.

There exist some works on beam alignment algorithms for reducing the latency of initial beam establishment. The hierarchical beam search algorithms define a wide beam codebook and a narrow beam codebook, respectively. Each wide beam direction contains a number of narrow beams. The BS first sweeps the wide beams and determines the best wide beam, then sweeps the narrow beams within the direction of best wide beam\cite{9134786,9918162}. The deep neural network (DNN) beam alignment algorithms can predict the best direction by measuring the RSRP from a subset of all available beams, and therefore, reduce the latency of initial beam establishment\cite{9513738,9653011,9690703}.

In the context of joint communications and sensing (JCAS), sensing-assisted BF can also be exploited to improve beam alignment using sensing results\cite{9540344}. In recent years, there has been an increasing amount of studies on this topic. But most studies have only focused on beam alignment under the assumption that a sensing object and a communication device are co-located. Therefore, by using the sensing results, e.g., the position of a vehicle, the BS can aim a narrow beam at the vehicle instead of finding the UE with the conventional feedback-based beam alignment method. For example, in \cite{9171304}, the authors proposed an extended Kalman filtering framework to sense the angular information of the vehicles. With this sensing information, the communication beam tracking overheads can be reduced. In \cite{9246715}, a Bayesian predictive BF scheme was proposed to improve the angle estimation of beam alignment. It has been verified that the sensing-assisted BF can applied to complicated roadway scenarios and improve the beam alignment accuracy\cite{10061429}. In \cite{9947033}, the authors exploited varying beamwidths to solve the beam misalignment because the sensing target and the UE cannot precisely coincide. However, the previous studies of sensing-assisted BF have not dealt with the high latency during initial beam establishment using narrow beams.

\begin{table}[t]
	\caption{ITU and Hexa-X permittivity and conductivity}
	\begin{center}
		\begin{tabular}{cccc}
			\hline
			Material & $\varepsilon_r$ & $\sigma_c$ & $\sigma_d$\\
			\hline
			Glass & 6.31 & 0.0036 & 1.3394\\
			Plaster & 2.73 & 0.0085 & 0.9395\\
			Plywood & 1.8 & 0.006 & 1 \\
			Glass wool & 1.2 & 0.002 & 1.2 \\
			Polystyrene & 1.05 & 0.000008 & 1.1 \\
			\hline
		\end{tabular}
	\end{center}
	\label{tab_1}
\end{table}
\begin{figure}[t]
	\centerline{\includegraphics[width=1\linewidth, height=10cm, keepaspectratio]{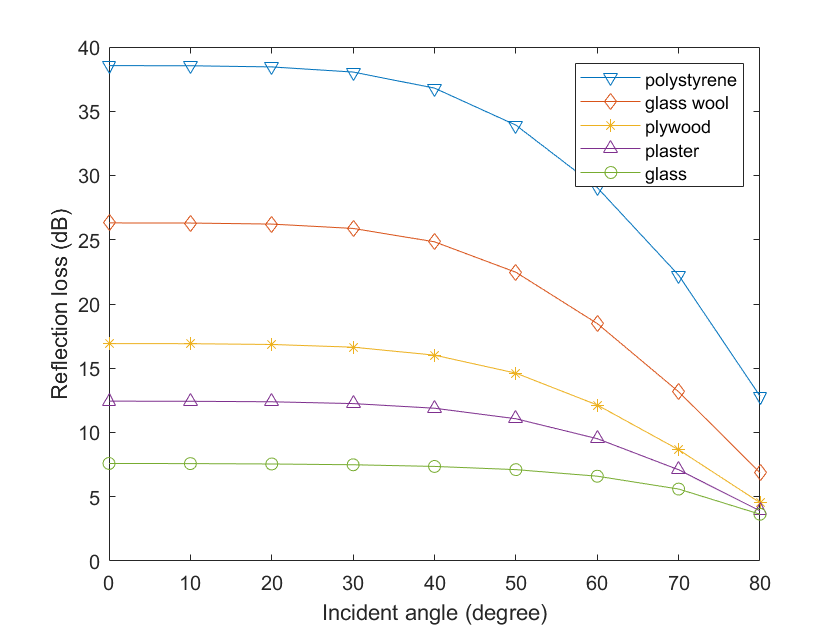}}
	\caption{The RLs induced by five common building materials at 100~GHz.}
	\label{fig_1}
\end{figure}

To tackle this problem, we propose a sensing-assisted BF method to identify the suitable beam direction with a much lower latency. The main idea of the proposed method is to identify the materials of the reflectors in the environment, thereby estimating the reflection losses (RL) of all beam directions according to the material information. The BS first scans the beam directions where low RL reflectors exist. Under these circumstances, there is a high probability that a trajectory with low overall path loss (PL) can be found quickly.

The paper is structured as follows. Section~II recaps the material sensing method using JCAS networks. Section~III proposes a material sensing-based initial beam establishment method for JCAS systems. In Section~IV, we evaluate the performance of the proposed method. In Section~V, the conclusion is drawn.

\section{RL-Based Material Identification}
Our previous studies in \cite{9482524,eurasip,9815822} have proposed a series of RL-based material identification methods for JCAS systems. The basic principle of these methods is described as follows. RL is determined by the material electrical properties (including permittivity and conductivity) and the incident angle. Fresnel reflection coefficient $r_{\text{TE}}$ for transverse electric (TE) polarization is given by
\begin{equation}\label{eqn_1}
	r_{\text{TE}}(\theta_\text{i}) = \frac{\text{cos}{\theta_\text{i}}-\sqrt{\eta-\text{sin}^2{\theta_\text{i}}}}{\text{cos}{\theta_\text{i}}+\sqrt{\eta-\text{sin}^2{\theta_\text{i}}}},
\end{equation}
where $\theta_\text{i}$ is the incident angle, $\eta$ is the relative permittivity of a specific material, and
\begin{equation}\label{eqn_2}
	\eta = \varepsilon_r -\text{j}\frac{17.98\sigma_cf^{\sigma_d}_\text{c}}{f_\text{c}},
\end{equation}
where $f_\text{c}$ is the carrier frequency, $\varepsilon_r$ is the dielectric relative permittivity, $\sigma_c$ and $\sigma_d$ are the material properties that determine the dielectric relative conductivity. Parameters $\varepsilon_r$, $\sigma_c$, and $\sigma_d$ of five common building materials have been measured by ITU\cite{itu} and Hexa-X\cite{d23}. These parameters are tabulated in \mbox{Table~I}. The RL for TE polarization can be calculated by\cite{d23}

\begin{equation}\label{eqn_3}
	RL(\theta_\text{i}) = 10\text{log}_{10}|\frac{1}{r_{\text{TE}}(\theta_\text{i})}|^2.
\end{equation}

The RLs induced by reflecting surfaces made of glass, plaster, plywood, glass wool, and polystyrene at incident angles from $0^{\circ}$ to $80^{\circ}$ are plotted in Fig.~1 and listed in Table~II. RL can also be obtained from measurements\cite{eurasip}. The overall PL of a trajectory in decibels is the difference between the transmitted power $P_{\text{TX}}$ at the transmitter (TX) side and the received power $P_{\text{RX}}$ at the receiver (RX) side,
\begin{equation}\label{eqn_4}
	PL = P_{\text{TX}} - P_{\text{RX}}.
\end{equation}

The overall PL of a single-bounce reflection trajectory with length $d$ can be expressed as the sum of free space path loss (FSPL) and RL,
\begin{equation}\label{eqn_5}
	PL(f_\text{c},d,\theta_\text{i}) = FSPL(f_\text{c},d) + RL(\theta_\text{i}).
\end{equation}

FSPL can be calculated by Friis' equation,
\begin{equation}\label{eqn_6}
	FSPL(f_\text{c},d) = 20\text{log}_{10}(d) + 20\text{log}_{10}(f_\text{c}) + 20\text{log}_{10}(\frac{4\pi}{c}),
\end{equation}
where $c$ is light speed.

From the measured $P_{\text{RX}}$ and trajectory information (length $d$ and incident angle $\theta_\text{i}$), the RL of a single-bounce reflection trajectory can be obtained. From Fig.~1, we can observe that different materials induce significantly different RLs at the same incident angle. Based on this fact, the material of a reflecting surface can be estimated by searching the RL at a specific incident angle in an RL database (e.g., the database in Table~II).

\begin{table*}[htb]
	\centering
	\caption{A database of RL induced by glass, plaster, plywood, glass wool, and polystyrene at 100 GHz}\label{tab_1}
	\begin{tabular}{@{} l*{9}{>{$}c<{$}} @{}}
		\toprule
		\textbf{Material} & \multicolumn{9}{c@{}}{\textbf{RL in decibel at different incident angle in degree}}\\
		\cmidrule(l){2-9}
		& 0^{\circ} &  10^{\circ} &  20^{\circ} &  30^{\circ} &  40^{\circ} &  50^{\circ} &  60^{\circ} &  70^{\circ} &  80^{\circ} \\
		\midrule
		Glass & 7.59 & 7.58 & 7.55 & 7.49 & 7.36 & 7.11 & 6.6 & 5.6 & 3.64 \\[1ex]
		
		Plaster & 12.44 & 12.43 & 12.39 & 12.25 & 11.89 & 11.07 & 9.52 & 7.1 & 3.9 \\[1ex]
		
		Plywood & 16.92 & 16.91 & 16.85 & 16.64 & 16.02 & 14.62 & 12.14 & 8.66 & 4.54 \\[1ex]
		
		Glass wool & 26.31 & 26.3 & 26.22 & 25.88 & 24.84 & 22.47 & 18.48 & 13.17 & 6.88 \\[1ex]
		
		Polystyrene & 38.55 & 38.54 & 38.45 & 38.05 & 36.8 & 33.92 & 29.06 & 22.24 & 12.83 \\[1ex]
		
	\end{tabular}
\end{table*}
\begin{figure*}[!t]
	\centering
	\subfloat[]{\includegraphics[width=1\columnwidth]{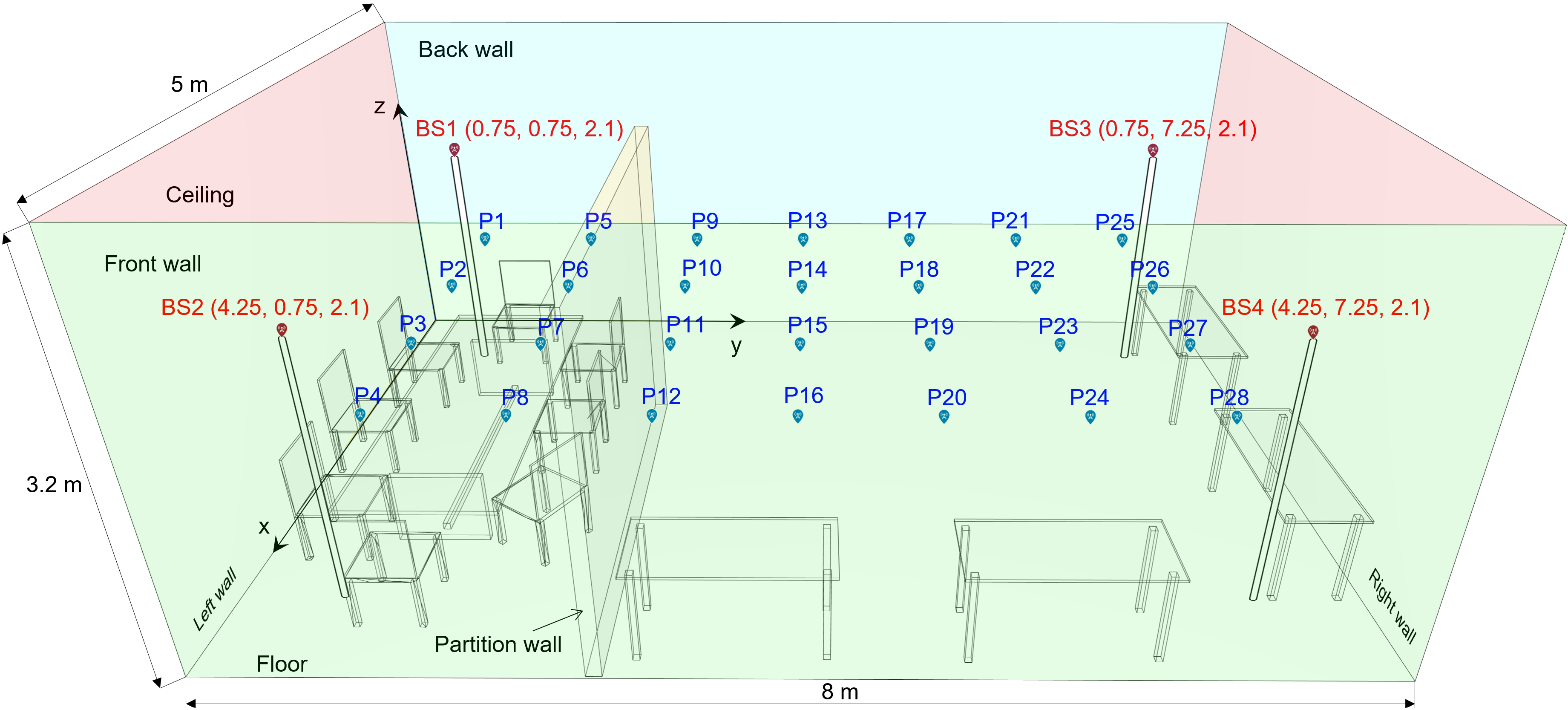}%
		\label{fig_first_case}}
	\hfil
	\subfloat[]{\includegraphics[width=0.85\columnwidth]{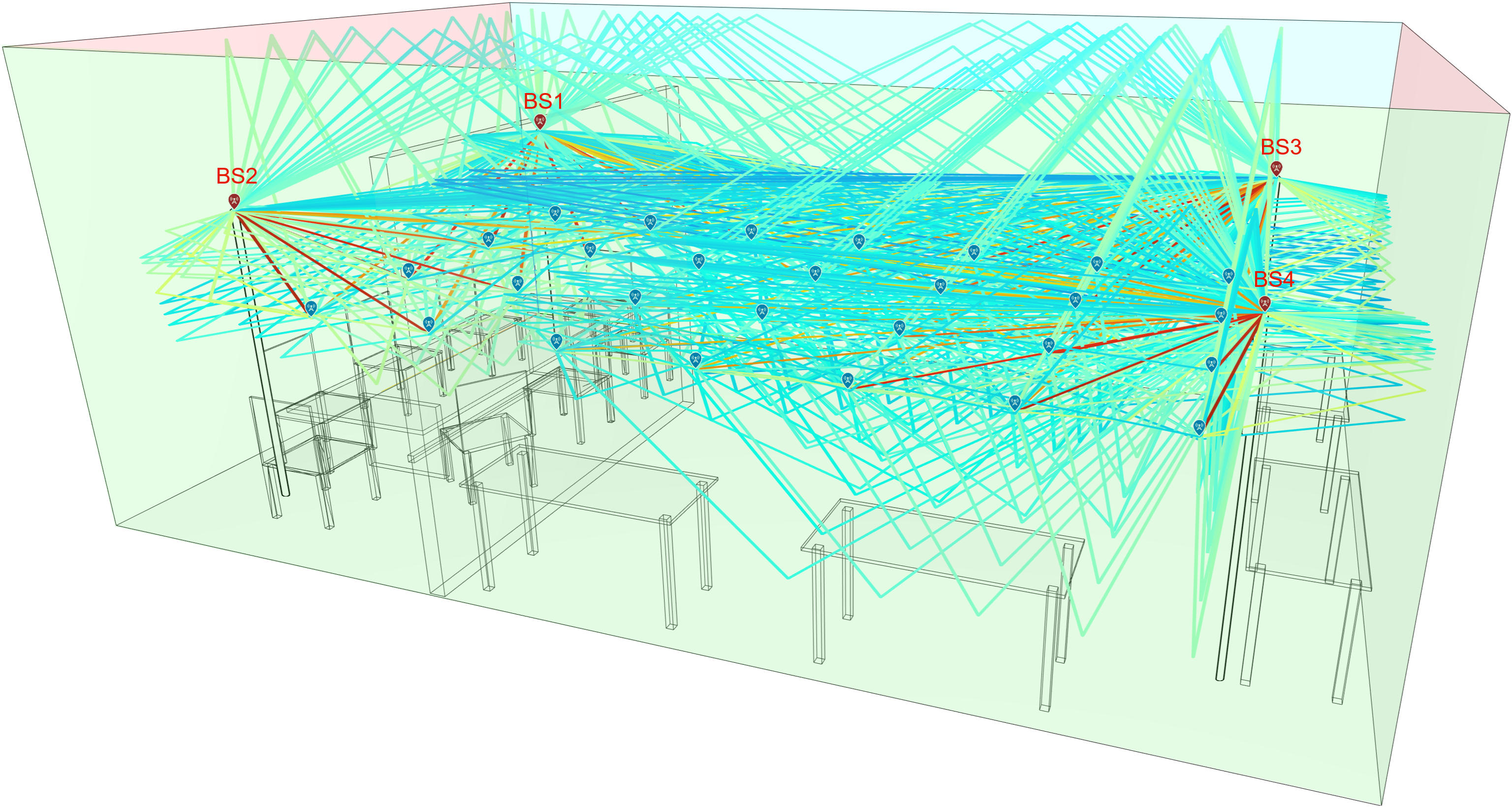}%
		\label{fig_second_case}}
	\caption{Illustration of (a) a simulated office with 4 BSs and 28 UE positions. (b) the trajectories between all the BS-UE position pairs calculated by ray tracing.}
	\label{fig_8}
\end{figure*}


\section{Material Sensing-Based Initial Beam Establishment Method}
In 5G NR, the BS exhaustively sweeps all the beam directions with a predefined scanning pattern, e.g., a sawtooth pattern, to search communication devices at unknown positions. However, this procedure becomes more time-consuming at mmWave and THz frequency bands due to the large volume of beam space. Moreover, with the help of the high gain of narrow beams, a reflecting trajectory between the TX and RX may still provide good connectivity. Thus, the high latency problem can be converted to another issue about how to find a reflecting trajectory with low overall PL quickly.

As mentioned in Section~II, the overall PL of a reflecting trajectory consists of FSPL and RL. When a BS searches a UE and sweeps the narrow beam to a specific beam direction, it is uncertain if a path exists between them. Even if the path exists, the path length, which can be transferred to FSPL, is entirely unpredictable before establishing the connection. However, the RL is predictable if the material information of the environment is available beforehand. As shown in Fig.~1, glass induces significantly lower RLs than the RLs induced by other materials. Therefore, the beam directions aiming at surfaces made of glass have a higher probability of finding and establishing a reflecting trajectory with lower overall PL than the beam directions aiming at other materials. Based on the above analysis, we propose a material sensing-assisted beam sweeping method with low latency during initial beam establishment. The method consists of the following steps:

\begin{itemize}
	\item Step 1: Import the three-dimensional (3D) map of the environment and identify the material information of objects in the environment using the methods proposed in \cite{9482524,eurasip,9815822}.
	\item Step 2: Prioritize the beam directions based on the materials of reflectors in the environment. The beam directions to material with lower RL have the higher priority during beam sweeping and vice versa. The priority of beam directions determines the beam sweeping order.
	\item Step 3: During the initial beam establishment phase, the BS will scan the beam directions and transmit SSBs in decreasing priority order. The UE measures and reports the RSRP of SSBs. Once the RSRP exceeds a specific threshold, the connection is established immediately without sweeping the rest of the beam directions.
	\item Step 4 (Optional): Because the trajectory established in step~3 may not be the best, beam update can be executed to search for the best trajectory, e.g., the line-of-sight (LOS) trajectory. 
\end{itemize}

\section{Simulation Results}
\addtolength{\topmargin}{0.03in}
\begin{figure}[!t]
	\centering
	\includegraphics[width=0.7\columnwidth]{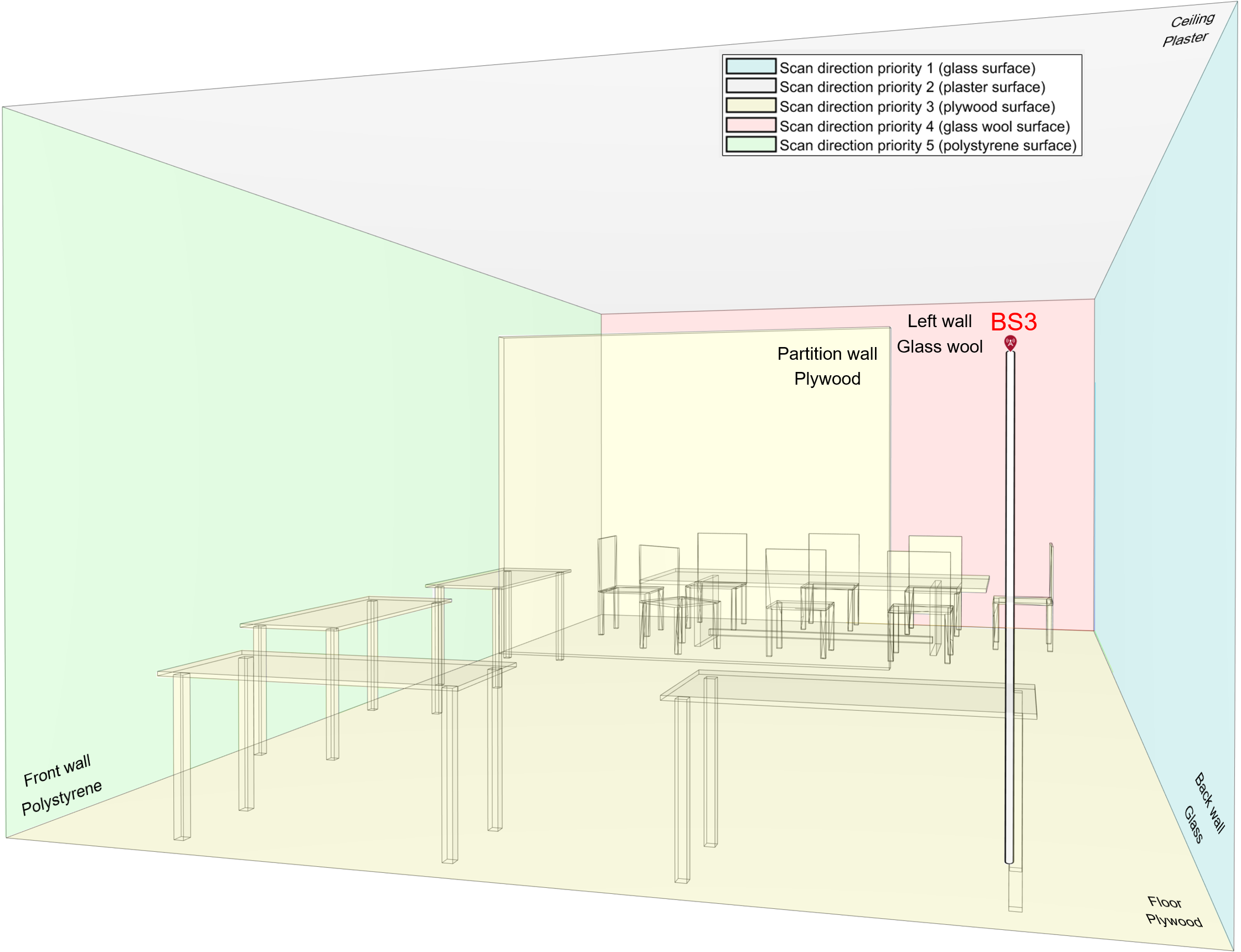}
	\caption{The perspective image of the office from a viewpoint behind BS3.}
	\label{fig_3}
\end{figure}

To validate the effectiveness of the proposed method, we present some MATLAB simulation results in this section. As shown in Fig.~2(a), a 3D map of an office with dimensions of 5~m$\times$8~m$\times$3.2~m in standard tessellation language (STL) format is imported. STL format is a 3D map format and can be created by various 3D modeling software. The materials of surfaces in the office are specified as follows: the front wall is polystyrene, the back wall is glass, the side wall is glass wool, the floor, tables, chairs, and the partition wall is plywood, and the ceiling is plaster. Different colors indicate the materials of surfaces. Four BSs (BS1-BS4) are mounted close to the upper corner of the office. To reduce simulation complexity, we assume a UE can only be located at one of 28 fixed positions (P1-P28). Then, ray tracing, a popular technique for predicting the radio propagation paths, is performed between each BS-UE position pair. Since trajectories with more than two reflection bounces usually cannot provide good connectivity, only single-bounce reflection is considered in the simulation. Fig.~2(b) illustrates all the 430 trajectories between each BS-UE position pair.

BS3 at (0.75, 7.25, 2.1) and UE position P14 at \mbox{(2, 4, 1.3)} are chosen as an example to show how the proposed method can reduce the latency of initial beam establishment. Fig.~3 depicts the perspective image of the office from a viewpoint behind BS3. According to the RL data in Fig.~1 and Table~II, the back wall made of glass induces the lowest RL. Therefore, BS3 should first scan the beam directions to the back wall (scan direction priority 1). If no trajectory is found, scan the beam directions to the ceiling (scan direction priority 2), and so on. For the antenna configuration in the simulation, the antenna observable region, in other words, the field of view (FOV), is set to $\pm70^{\circ}$ both in azimuth and elevation. The half-power beamwidth is set to $5^{\circ}$. To boresight the antenna array to the office's center, the antenna array's mechanical orientation is set to $7.7^{\circ}$ down-tilt and $28.3^{\circ}$ bearing. Fig.~4 illustrates the five single-bounce reflection trajectories between BS3 and P14 obtained by ray tracing. The information about the five trajectories, including the reflector's material, path length, FSPL, incident angle, RL, and overall PL, is tabulated in Table~III. As seen from Table~III, trajectory~1 has much lower RL than the RLs of other trajectories. Although the FSPL of trajectory~1 is not the lowest, trajectory~1 still has the lowest overall PL among all trajectories. If only one non-LOS trajectory between BS3 and a UE at P14 can be established, trajectory~1 is the best option.
\begin{table*}[t]
	\caption{The information of trajectories between BS3 and P14, BS4 and P16}
	\begin{center}
		\begin{tabular}{cccccccccc}
			\hline
			BS-position pair & Trajectory & Reflector & Material & Path length (m) & FSPL (dB) & Incident angle & RL (dB) & Overall PL (dB)\\
			
			\hline
			& 1 & back wall & glass & 4.6 & 85.71 & 50.7 & 7.08 & 92.79\\
			& 2 & front wall & polystyrene & 8 & 90.51 & 24.9 & 38.31 & 128.82\\
			BS3-P14 & 3 & ceiling & plaster & 4.43 & 85.16 & 49.3 & 11.14 & 96.3\\
			& 4 & floor & plywood & 4.87 & 86.2 & 45.6 & 15.36 & 101.56\\
			& 5 & partition wall & plywood & 5.97 & 87.97 & 14.3 & 16.9 & 104.87\\
			\hline
			& 6 & back wall & glass & 8.88 & 91.42 & 22.1 & 7.54 & 98.96\\
			& 7 & front wall & polystyrene & 3.79 & 84.02 & 62.3 & 27.67 & 111.69\\
			BS4-P16 & 8 & ceiling & plaster & 4.43 & 85.38 & 47.4 & 11.34 & 96.72\\
			& 9 & floor & plywood & 4.71 & 85.91 & 43.8 & 15.61 & 101.52\\
			& 10 & partition wall & plywood & 5.85 & 87.79 & 12 & 16.91 & 104.7\\
			\hline
		\end{tabular}
	\end{center}
	\label{tab_3}
\end{table*}
\begin{figure}[t]
	\centering
	\includegraphics[width=0.7\columnwidth]{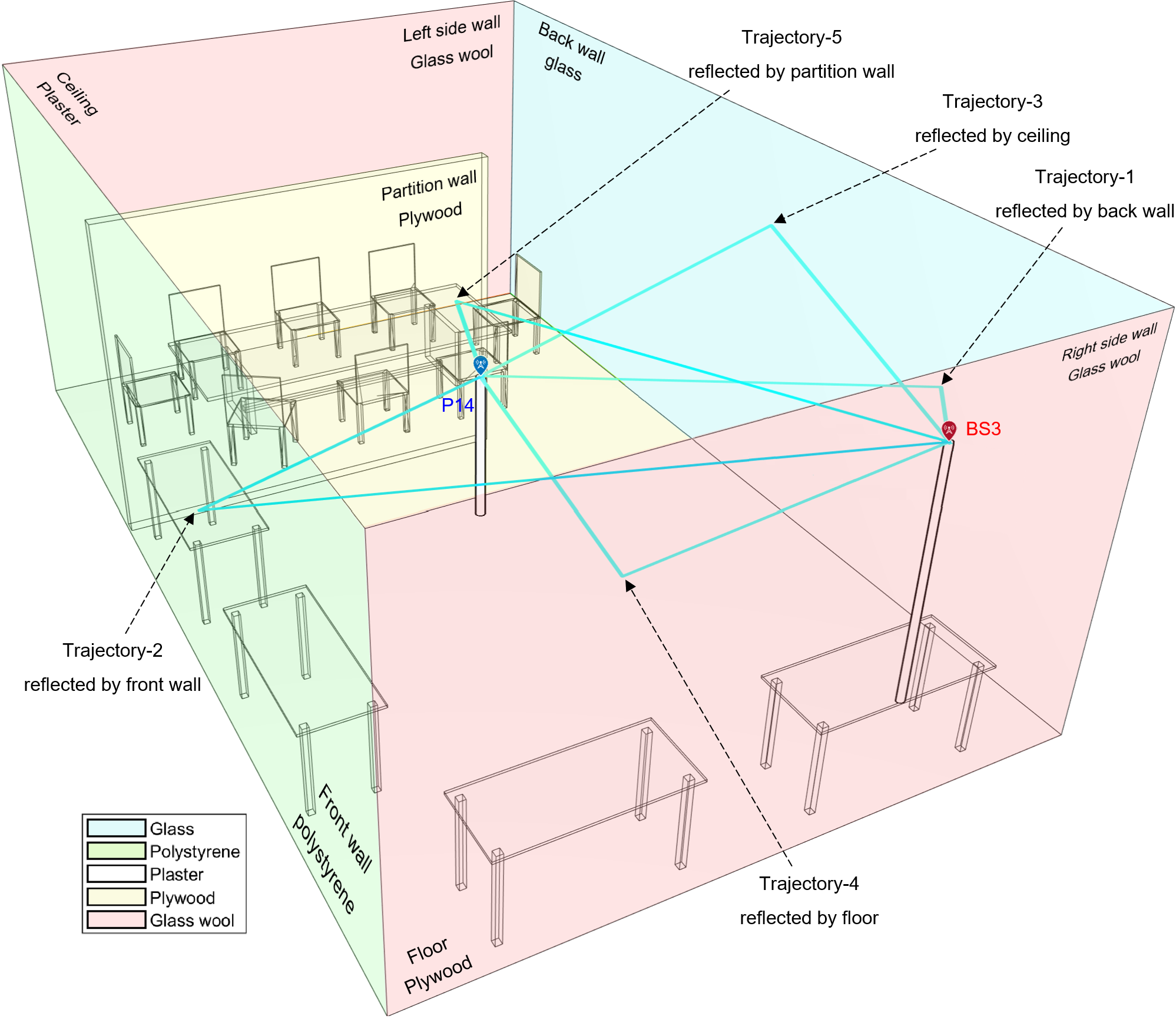}
	\caption{Five single-bounce reflection trajectories between BS3 and P14.}
	\label{fig_4}
\end{figure}
\begin{figure}[t]
	\centering
	\includegraphics[width=0.8\columnwidth]{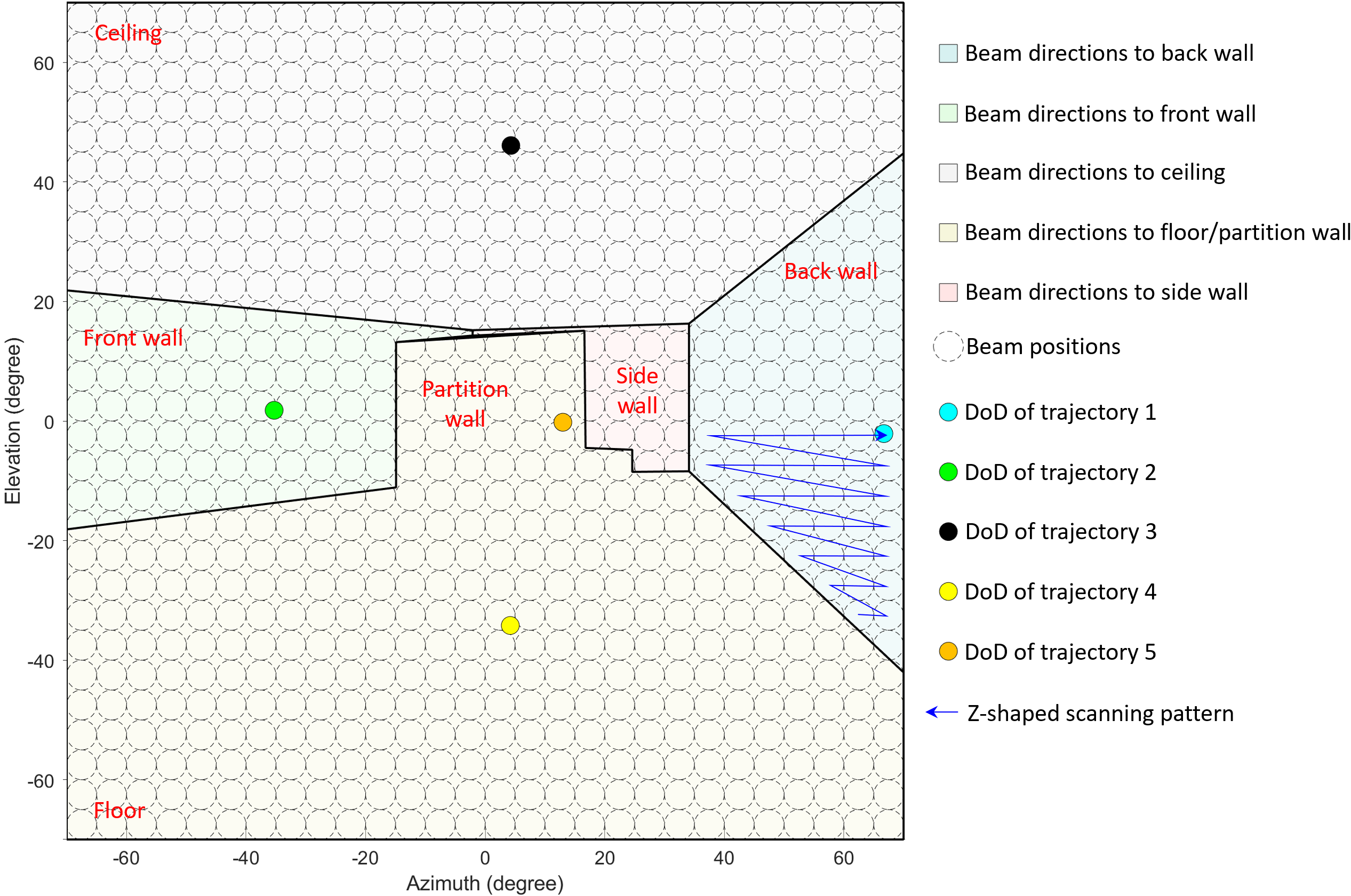}
	\caption{The view and the DoD of trajectories 1-5 between BS3 and P14, seen from BS3.}
	\label{fig_5}
\end{figure}

\begin{figure*}[!t]
	\centering
	\includegraphics[width=2\columnwidth]{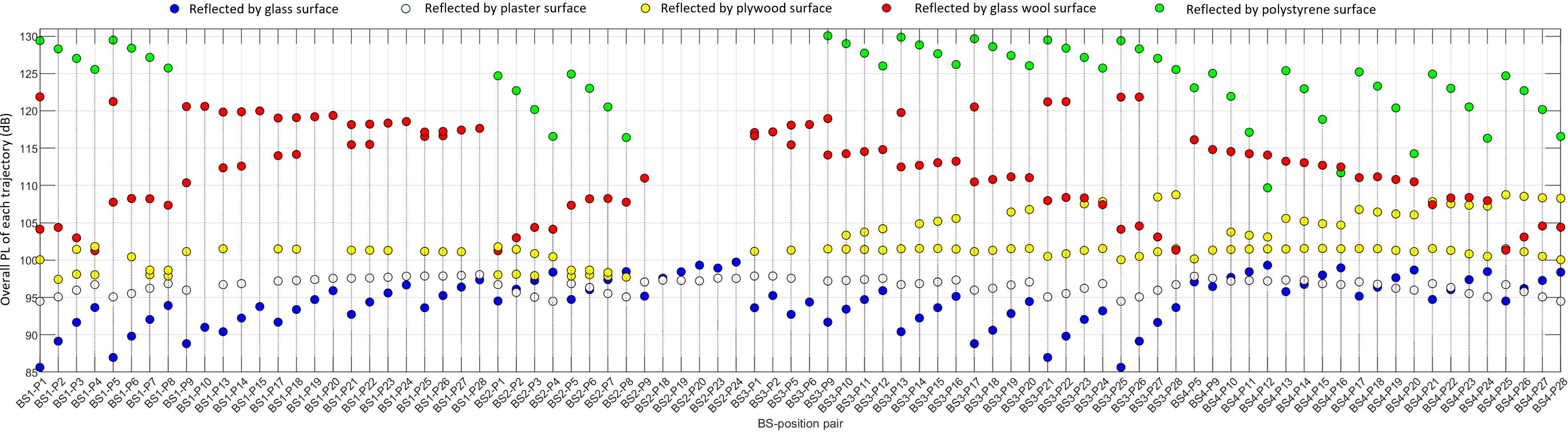}
	\caption{The RLs of the 430 trajectories illustrated in Fig.~2(b).}
	\label{fig_3}
\end{figure*}

\begin{figure}[t]
	\centering
	\includegraphics[width=0.7\columnwidth]{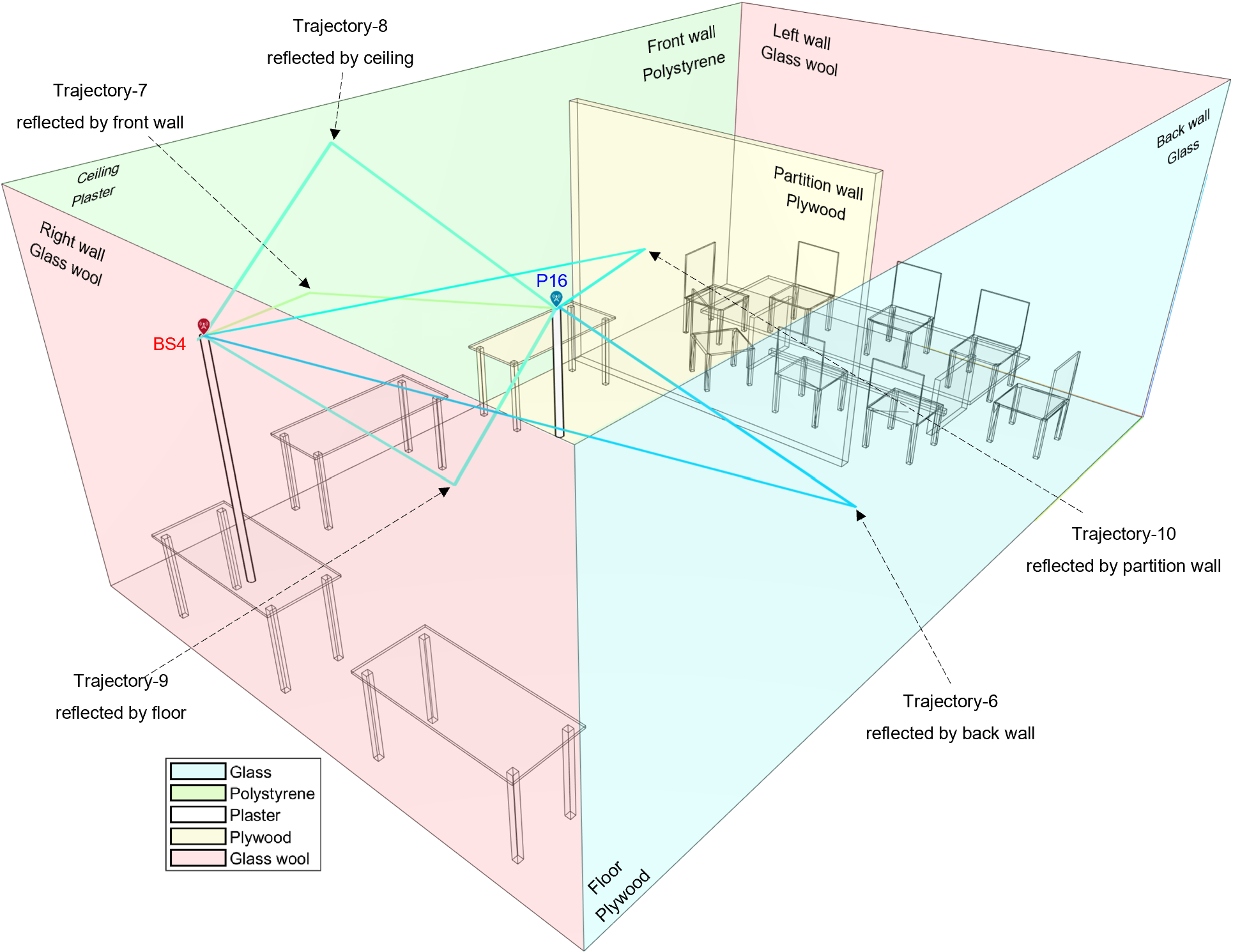}
	\caption{Five single-bounce reflection trajectories between BS4 and P16.}
	\label{fig_7}
\end{figure}

Fig.~5 shows what BS3 could ``see" in the office within its FOV. The cyan/green/black/yellow/orange markers ``$\bigcirc$" indicate the directions of departure (DoD) of trajectories 1/2/3/4/5 illustrated in Fig.~4. For example, a beam aiming at the cyan marker will be reflected by the back wall and received by the UE at P14, i.e., trajectory~1 will be established between BS3 and UE at P14. The dashed circles in Fig.~5 represent the beam positions that BS3 can aim at. There are 784 beam directions, indicating that the exhaustive beam search procedure must scan 784 beam directions before establishing the connection. With the proposed method, BS3 first scans the glass back wall. We assume that the BS3 scans the back wall in a sawtooth pattern. In other words, BS3 sweeps the beam horizontally from left to right in the bottom and then steps up in elevation, as the blue polygonal arrow indicated in Fig.~5. After 34 sweeps, trajectory~1 will be established. The remaining 750 beam positions need not be scanned any longer. Compared to the exhaustive beam search, the latency of initial beam establishment of the proposed method is reduced by 95.7\% (750/784) in this case.


To assess if the proposed method can also achieve similar performance in different cases, we perform the proposed method for all BS-UE position pairs and evaluate the performance. Fig.~6 provides the overall RLs of all the trajectories illustrated in Fig.~2(b). The x-axis shows all the BS-UE position pairs from BS1-P1 to BS4-P28. The y-axis shows the overall PLs of the trajectories. The RLs of trajectories reflected by glass/plaster/plywood/glass wool/polystyrene are illustrated by blue/white/yellow/red/green dots. We consider the trajectory with the lowest overall PL as the best trajectory. From Fig.~6, we can see that there is a high probability that the trajectory reflected by the glass surface is most likely the best. The total number of BS-UE position pairs is 84, out of which 73.8\% of the pairs (i.e., 62 pairs) have the best trajectory reflected by glass surface, and the remaining 26.6\% of the pairs (i.e., 22 pairs) have the best trajectory reflected by plaster surface. Hence, the proposed method can achieve 73.8\% accuracy in finding the best trajectory by first scanning the glass surface. An explanation for the failed 22 pairs is that the trajectory reflected by the glass surface is much longer than that reflected by the plaster surface. The lower RL induced by the glass cannot counteract the large FSPL due to the long trajectory. Take the trajectories between BS4 and P16 as examples. Table~III provides data for all the trajectories between BS4 and P16. As illustrated in Fig.~7, BS4 and P16 are far from the glassy back wall. Trajectory~6 reflected by the glass is 8.88~m in length, much longer than trajectory~8 reflected by plaster with a length of 4.43~m. The FSPL of trajectory~6 (91.42~dB) is 6.04~dB higher than that of trajectory~8 (85.38~dB). Therefore, trajectory~6 has a higher overall PL than trajectory~8, although trajectory~6 has a lower RL. However, trajectory~6 is still considered a good option for BS4-P16 pair compared to the trajectories reflected by plywood, glass wool, and polystyrene.

To assess the latency performance of the proposed method, the establishment latencies of all trajectories reflected by the back wall are analyzed. Figs.~8(a)-(d) show the views from BS1 to BS4, respectively. The cyan markers indicate the DoD of trajectories reflected by the back wall. Using the sawtooth scanning pattern, the mean number of beam sweeps to find these trajectories is 36.4, indicating that the average latency of initial beam establishment of the proposed method is reduced by 95.4\% compared to the exhaustive beam search. In addition, all the reflection points are located in the midst of the back wall, as seen from Figs.~8(a)-(d). This is because the simulation restricts the height of the UEs to 1.3~m, representing a realistic value for portable communication devices. This constraint increases the chance of reflection points located in the midst of the walls. This finding suggests that the latency in the simulation would be reduced further by first scanning the midst of the back wall. Therefore, the use of the distribution of communication devices and designing new scanning patterns could be the means of reducing latency.

The proposed method can be enhanced further by combining it with existing solutions such as hierarchical beam search algorithms and DNN-based beam search algorithms. For example, in Fig.~5, the narrow beams pointing to the back wall can be grouped into wider beams. Then, the hierarchical beam search algorithms can be implemented with the proposed method. DNN-based beam search algorithms can also be used to reduce the latency by measuring the RSRP from a subset of all narrow beams pointing to the back wall.

\begin{figure*}[!t]
	\centering
	\subfloat[]{\includegraphics[width=0.5\columnwidth]{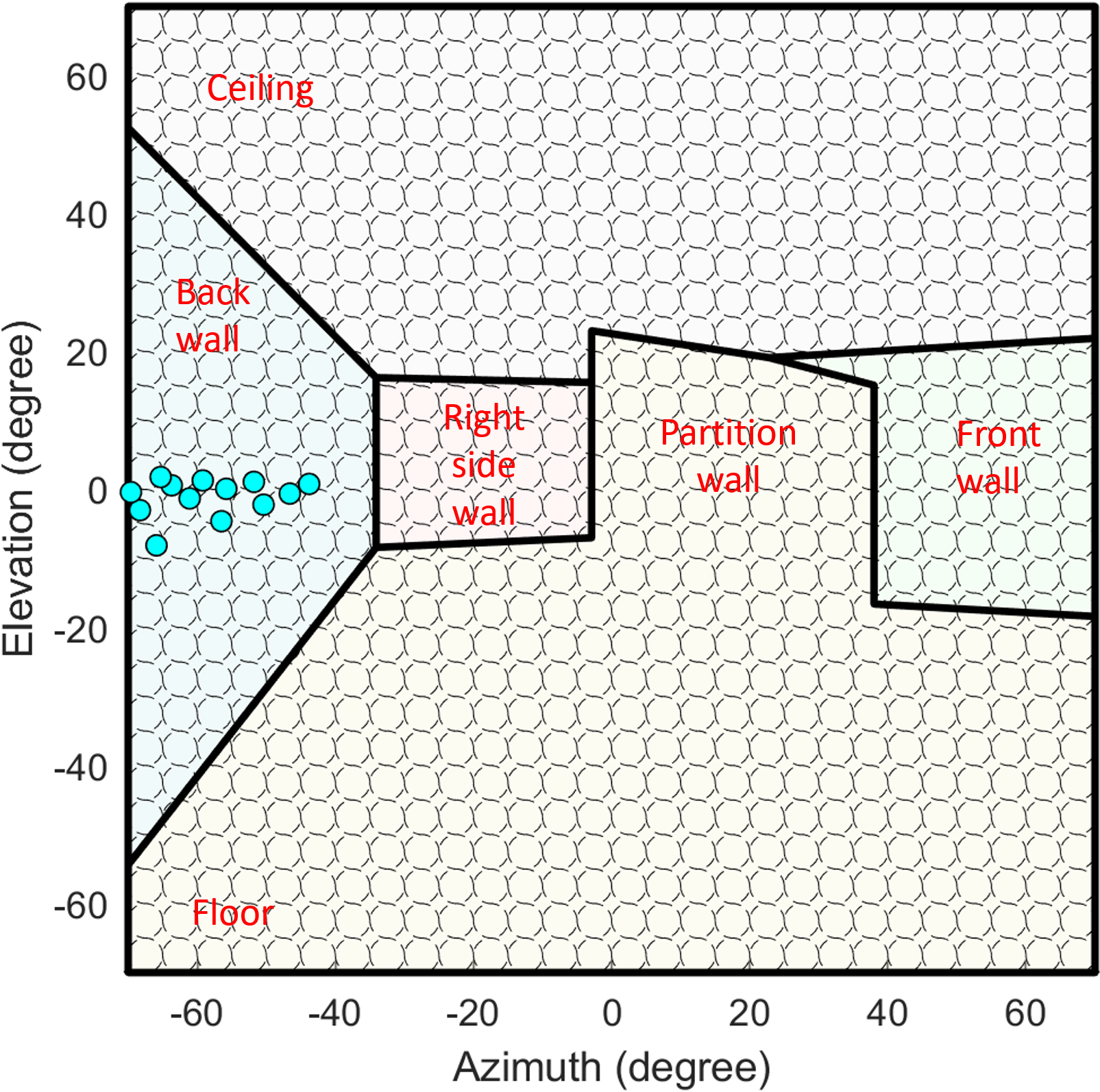}%
		\label{fig_first_case}}
	\hfil
	\subfloat[]{\includegraphics[width=0.5\columnwidth]{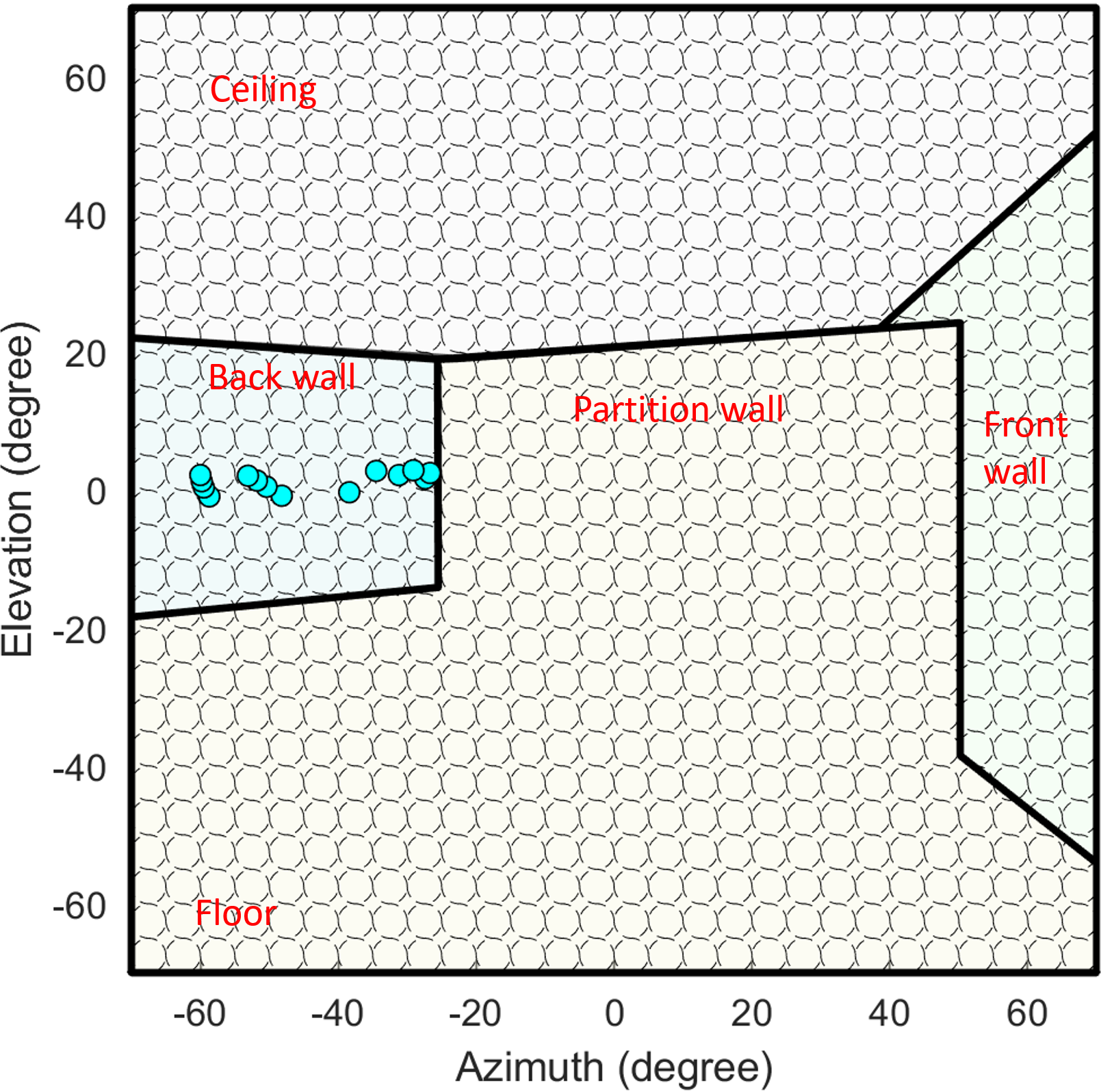}%
		\label{fig_second_case}}
	\hfil
	\subfloat[]{\includegraphics[width=0.5\columnwidth]{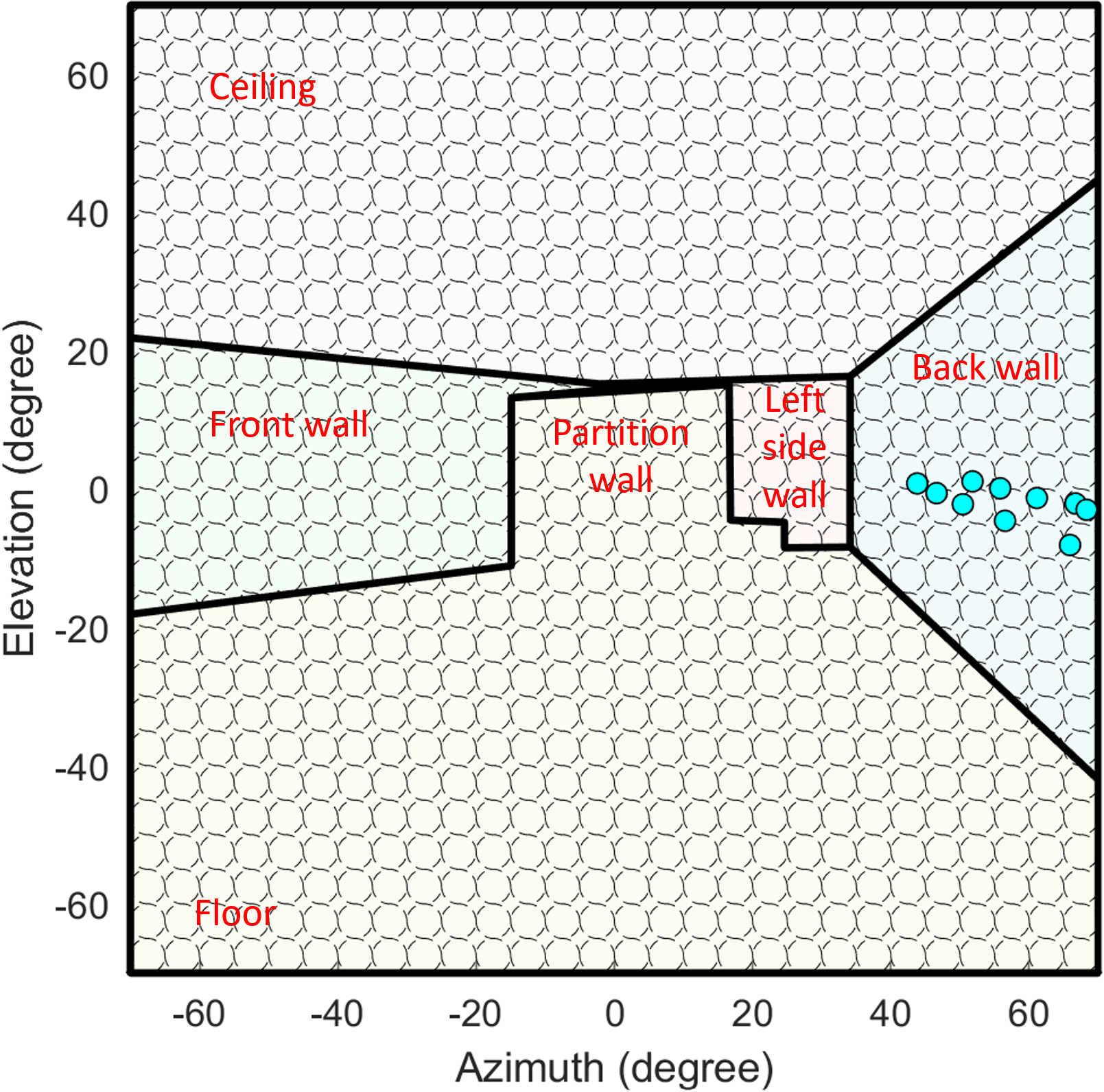}%
		\label{fig_second_case}}
	\hfil
	\subfloat[]{\includegraphics[width=0.5\columnwidth]{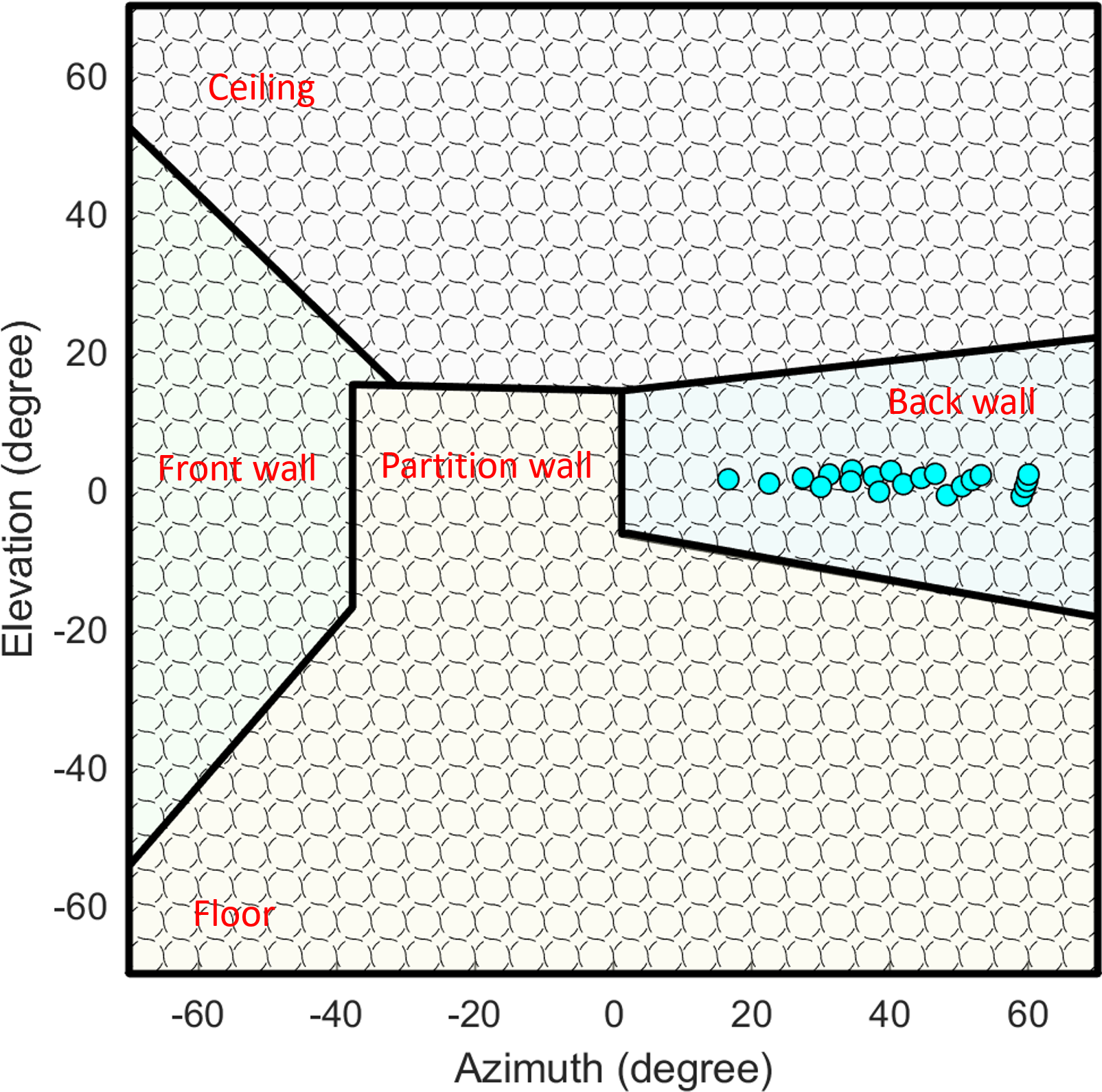}%
		\label{fig_second_case}}
	\caption{The views and the DoD of trajectories reflected by the back wall seen from (a) BS1 (b) BS2 (c) BS3 (d) BS4.}
	\label{fig_8}
\end{figure*}
\section{Conclusions}
Sensing-assisted communications is one of the promising techniques for JCAS systems. In this paper, we have proposed a material sensing-assisted BF method to reduce the latency of initial beam establishment for high frequency bands. The RL at different beam directions can be estimated with pre-identified material information in the environment. Then, by scanning the beam directions with low RL first, the BS may find and establish a suitable beam pair without scanning all the beam directions, thereby significantly reducing the latency of initial beam establishment. Furthermore, the proposed method can be combined with the hierarchical beam search algorithms and DNN-based beam search algorithms to reduce latency further. This material sensing-assisted BF method can be a technology that will enable 6G networks to satisfy the stringent low latency requirements in some ultra-reliable low-latency communications use cases.

\section{Acknowledgment}
This work has been funded by the National Key Research and Development Program of China (No.2022YFB2902405).

\bibliographystyle{IEEEtran}
\bibliography{ref} 

\color{red}
\end{document}